\documentclass[aps,prd,twocolumn,showpacs,superscriptaddress,groupedaddress]{revtex4}
\usepackage[latin1]{inputenc}
\usepackage{graphicx}
\usepackage{amssymb}
\usepackage{color}
\usepackage{float}
\usepackage{amsmath}
\usepackage{amsfonts}
\usepackage{dcolumn}
\usepackage{hyperref}
\usepackage{amsthm}
\usepackage{color}

\def\3nab{\tilde{\nabla}}

\def\be {\begin{equation}}
\def\ee {\end{equation}}
\def\ba {\begin{eqnarray}}
\def\ea {\end{eqnarray}}

\newcommand{\E}{{\mathcal E}}

\newcommand{\barray}{\begin{array}}
\newcommand{\earray}{\end{array}}

\begin{document}

\title{From a collapsing radiating star to an evaporating black hole:  A smooth transition from classical to quantum entropy}
\author{Sarbari Guha}
\email{guha@sxccal.edu}
\affiliation{Department of Physics, St. Xavier's College (Autonomous), 30 Mother Teresa Sarani, Kolkata, India}
\author{Shamima Khan}
\email{shmimakhan094@gmail.com}
\affiliation{Department of Physics, St. Xavier's College (Autonomous), 30 Mother Teresa Sarani, Kolkata, India}
\author{Rituparno Goswami}
\email{goswami@ukzn.ac.za}
\affiliation{Astrophysics Research Centre, School of Mathematics, Statistics and Computer Science, University of KwaZulu-Natal, Private Bag X54001, Durban 4000, South Africa}

 \begin{abstract}
We present a robust mechanism, where the geometrical free-gravitational entropy of an isolated astrophysical radiating star undergoing continual gravitational collapse, as measured by an external observer, makes a smooth transition to the Bekenstein-Hawking entropy at the onset of the horizon formation and in the late times of black hole evaporation. It is interesting to note that both in the classical regime and the semi classical evaporating black hole regime, the matter is radiated via the Vaidya exterior surrounding the radiating star, as well as the evaporating black hole. Our result, being independent of the interior matter dynamics of the collapsing star, clearly indicates that the Bekenstein-Hawking entropy and its non-extensive nature indeed originates from the Riemannian geometry, which dictates the free-gravity entropy in general relativity.
\end{abstract}

\maketitle

Black holes (BHs) are unique and extreme astrophysical objects that were first envisioned by Laplace \cite{Laplace,HawkingEllis}, who noted that the ``{\it attractive force of a heavenly body can be so large that even light cannot flow out of it} ''. This was later put on a solid theoretical footing by the predictions of General Relativity (GR) via the singularity theorems of Hawking, Penrose and Geroch \cite{HawkingEllis}. Recent precision observations by the Event Horizon Telescope (EHT) (see \cite{EHT} and related references), established the existence of BHs beyond reasonable doubt. These extreme objects lie on the boundary of classical and quantum descriptions of the universe and hence provide us with an excellent tool to probe the fundamental principles of gravitational physics.

The foundations of BH thermodynamics for an isolated and eternal black hole were laid by the pioneering works of Bekenstein \cite{Bekenstein} and Bardeen et al \cite{BCH}, and subsequently developed by several others \cite{HawkingEllis, Hawking, GibbonsHawking}, which have been discussed in detail in \cite{Page}. The extraordinary fact that BH entropy is a non-extensive property (that scales as the surface area and not the volume of the black hole), emerged naturally from rigorous quantum field theoretic (QFT) calculations on curved background. However, the important point here is, that BHs are not eternal, and they are dynamically produced by the continual gravitational collapse of massive stars at their final stages.

Therefore, the key question here is as follows: ``{\it Is it possible to define a geometrical mechanism that allows a smooth transition of classical gravitational entropy of a collapsing astrophysical star to the non-extensive quantum entropy of an evaporating black hole?}" This question is important as it would give us a solid geometrical foundation for the non-extensivity of BH entropy and shed new light on the transition process from the classical to quantum regime. Although \cite{Ellis:2014jja,DadGos} discussed in detail about the actual site of the emergence of Hawking radiation from accreting BH, this specific question about isolated and non-accreting BH was left unanswered.

The important hurdle that impeded a satisfactory answer to the question posed above, was the absence of any satisfactory energy momentum tensor for free gravity (gravity in the absence of matter). In his well-known Weyl Curvature Hypothesis (WCH), Penrose \cite{Penrose1,Penrose2} suggested that any definition of entropy of free gravitational fields (gravitational entropy), should incorporate the Weyl curvature tensor. Noting that the Bel-Robinson tensor \cite{bell1,Bel,bell2,robin},  $T_{abcd}\equiv\frac14\left(C_{eabf}C^{e~~~f}_{~cd}+C^*_{eabf}C^{*e~~~f}_{~cd}\right)$, is a unique Maxwellian tensor with a dimension  $L^{-4}$, that can be constructed from the Weyl tensor so as to act as the `super energy momentum' tensor for gravitational fields \cite{Maartens:1997fg,Sen1}, it was proposed that the symmetric 2-index square root $ \mathcal{T}_{ab}$ of the Bel-Robinson tensor (that uniquely exists for spacetimes that are Petrov type D and N) would act as the effective energy momentum tensor for free gravitational field \cite{CET}.

In this letter, we use the symmetric 2-index square root of the Bel-Robinson tensor, $ \mathcal{T}_{ab}$, as the energy momentum tensor for source-free gravitational field, to develop the novel geometrical mechanism, that is robust and answers the above question comprehensively. We show transparently, that the non-extensive property of the Bekenstein-Hawking entropy does originate from classical geometry, via the geometric interpretation of the free gravity entropy and this contributes to a smooth thermodynamic transition from the classical collapsing star to the evaporating BH.

To give a clear geometrical picture of the entire process, we consider a spherically symmetric configuration (which is Locally Rotationally Symmetric (LRS) class II and also a subclass of Petrov type D) with a timelike vector
$u^a$ (defined along the fluid flow lines) and $e^a$ being the unit vector along the preferred spatial direction of LRS spacetimes (orthogonal to $u^a$). It is important to note that although we consider the spherically symmetric background, the results obtained here are stable with respect to any non-spherical perturbations of the system upto the linear order (as we will discuss later).
The spacetime can be decomposed as \cite{ClarksonBarrett, BetschartClarkson,Ellis:2014jja}
\be\label{decomp}
g_{ab}=-u_au_b+e_ae_b+N_{ab} ,
\ee
where $N_{ab}$ is the 2-dimensional metric that spans the spherical 2-shells.

The geometry for the $u^a$ congruence are defined by the expansion scalar  $\Theta$, acceleration 3-vector $\dot{u}^a$ and the shear 3-tensor $\sigma_{ab}$. The timelike congruence naturally gives the electric part of the Weyl tensor, which is responsible for tidal forces and inhomogeneity, as $E_{ab}= C_{acbd}u^cu^d$, while the magnetic part $H_{ab}=C^*_{acbd}u^cu^d$,  vanishes due to the spherical symmetry. Using $u^a$, we decompose the energy momentum tensor of the matter to give the energy density $\mu$, isotropic pressure $p$, heat flux 3-vector $q^a$ and anisotropic stress 3-tensor $\pi_{ab}$. The only non-vanishing geometrical quantity related to the preferred spacelike congruence is the volume expansion $\phi$. Using this congruence, covariant scalars can be extracted from the above mentioned 3-vectors and 3-tensors as $\mathcal{A}=\dot{u}^ae_a$, $\Sigma=\sigma_{ab}e^ae^b$, $\mathcal{E}=E_{ab}e^ae^b$, $Q=q^ae_a$ and $\Pi=\pi_{ab}e^ae^b$.
Therefore, the set of covariant scalar quantities that fully describe the spherically symmetric class of spacetimes are
\be\label{set1}
\mathcal{D}\equiv\left\{\Theta, \mathcal{A}, \Sigma, \mathcal{E}, \phi, \mu, p, \Pi, Q\right\}\;.
\ee
These geometrical and thermodynamical scalars, together with their directional derivatives along $u^a$ (denoted by a dot) and projected directional derivative along $e^a$ (denoted by a hat) completely specify the corresponding Ricci identities and the Bianchi identities, and thereby specify the complete dynamics.

\begin{figure}[h!!!]
\begin{center}
\includegraphics[width=0.45\textwidth]{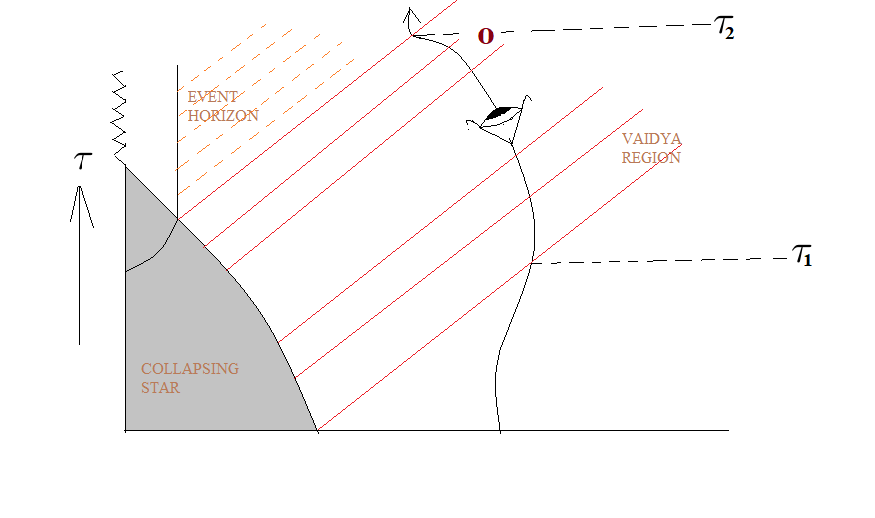}
\caption{A schematic diagram of the collapse of a radiating star as viewed by an external observer `O'. Solid lines denote classical Vaidya radiation whereas broken lines denote quantum radiation from the BH horizon.}\label{collapse}
\end{center}
\end{figure}

Let us now consider a collapsing and radiating spherical star, with thermal flux in the interior, which manifests itself as an outgoing Vaidya radiation (see figure \ref{collapse}, for a pictorial representation of the entire system). Hence, the interior spacetime is matched at the boundary $\mathcal{B}$ of the collapsing star to an exterior Vaidya geometry given by
\be\label{extmetric_2}
ds^{2}=-\left(1-\frac{2m(u)}{r}\right)du^{2}-2dudr+r^{2}d\Omega_2^2.
\ee
It is very important to note here that the Vaidya mass function $m(u)$ is not arbitrary, but entirely specified by the dynamics of the collapsing star via the boundary conditions at $\mathcal{B}$, as determined by Israel-Darmois matching conditions \cite{Darmois,Israel,Santos}, namely
\ba\label{matching}
m(u)\vert_\mathcal{B}&= &\mathcal{M}_I\vert_\mathcal{B,}\\
m(u)_{,u}\vert_\mathcal{B}&=&-\frac{1}{2K}(1-2\mathcal{M}\sqrt{K})(p_I+\Pi_I)\vert_\mathcal{B},
\ea
where $\mathcal{M}_I$, $p_{I}$ and $\Pi_I$ are the Misner Sharp mass and the interior pressure and anisotropic stress of the collapsing star respectively, and $K$ is the Gaussian curvature of the boundary shell. The second condition is the pressure balance condition at the boundary \cite{Santos}. Thus the Cauchy initial data (the value of $m(u)$ as well as its derivative $m(u)_{,u}$) at the boundary of the star, that is generated via the interior stellar dynamics, is sufficient for us to determine the Vaidya mass function for the entire Vaidya region by solving the Einstein field equations.

Now, for a timelike comoving observer in the Vaidya region, we have (where $F\equiv(1-{2m(u)}/{r})$)
\be\label{ue}
u^a = \frac{1}{\sqrt{F}}\frac{\partial}{\partial u}\;,\; \textrm{and} \;\; e^a=\frac{1}{\sqrt{F}}\frac{\partial}{\partial u}+\sqrt{F}\frac{\partial}{\partial r}\, ,
\ee
which immediately gives, from the field equations, the following quantities
\begin{eqnarray}
  \phi &=& \frac{2}{r}\sqrt{1- \frac{2m(u)}{r}} \, , \; \E =- \frac{2 m(u)}{r^3}\,, \\
  \Theta &=& \frac{m(u)_{, u}}{r}\frac{1}{ \left( \sqrt{1 - \frac{2m(u)}{r}} \right)^3}=\frac{3}{2} \Sigma \, , \\
  \mathcal{A} &=& \frac{m(u)}{r^2}\frac{1}{\sqrt{1-\frac{2m(u)}{r}}} - \Theta \, , \\
  \mu &=& -\frac{2 m(u)_{, u}}{r^2}\frac{1}{ \left( 1 - \frac{2m(u)}{r} \right) } = 3p = \frac{3}{2} \Pi = Q \, .
\end{eqnarray}
It is quite natural that all the geometrical and thermodynamical quantities for the Vaidya observer, depend on the Vaidya mass function and its derivative, which in turn is governed by the interior dynamics of the collapsing star.

The local free-gravity energy density, as seen by the Vaidya observer, is given by $ \mathcal{T}_{ab}u^au^b\equiv(1/4\pi)\vert \mathcal{E}\vert$ (see \cite{CET,Acquaviva,Goswami:2018zfo}), where  $ \mathcal{T}_{ab}$ is the unique 2-indexed square root of the Bell-Robinson tensor in the Vaidya region, given as \cite{Goswami:2018zfo}
\be\label{TfLRS}
\mathcal{T}_{ab}=\left[\frac32\vert\E\vert\left(u_au_b-e_ae_b\right)+\frac12\vert\E\vert g_{ab}\right]\;.
\ee
The causal temperature function of the free-gravity is the kinematic quantity related to the {\it extrinsic curvature of the 3-space orthogonal to the timelike congruence $u^a$}, and is given by the contraction of the covariant derivative of $u^a$ with the local future null cone \cite{CET,Acquaviva}). Thus we have $T=(1/\pi)\vert k^al^b\nabla_au_b\vert$, where $k^a\equiv1/(\sqrt{2})(u^a+e^a)$ and $l^a\equiv1/(\sqrt{2})(u^a-e^a)$ are future directed outgoing and ingoing null geodesics respectively. In terms of the geometrical variables, we get the local temperature function of the free gravity as
\be
T=\frac{1}{2\pi}\vert \mathcal{A}+(1/3)\Theta+\Sigma \vert\;.
\ee
The {\it local free-gravity entropy density} `$\zeta_G$' for the Vaidya observer then becomes
\be\label{entdens}
\zeta_G=\frac12\left[\frac{\vert \mathcal{E}\vert}{\vert\mathcal{A}+(1/3)\Theta+\Sigma\vert}\right]=\frac{1}{2r}\sqrt{1-\frac{2m(u)}{r}}\,.
\ee
Again, as expected, the interior dynamics of the collapsing star, directly affect the local free-gravity entropy density of the Vaidya observer. To find the total free-gravity entropy `$\mathcal{S}_G$'  as measured by the Vaidya observer, on the 3-space perpendicular to $u^a$, at any epoch, we perform a volume integration over the {\it proper 3-volume} (given by the 3-metric $h_{ab}=g_{ab}+u_au_b$) of the Vaidya region (from the boundary of the star to infinity), to get
\be\label{ent}
\mathcal{S}_G=\int_{r_\mathcal{B}}^{r_\infty}\zeta_G\sqrt{h}\,d^3x=2\pi\int_{r_\mathcal{B}}^{r_\infty}r dr .
\ee
The above equation reveals a couple of extraordinary aspects of free-gravity entropy exterior to a radiating and collapsing star:
\begin{enumerate}
\item The free-gravity entropy at any epoch is completely independent of the interior dynamics of the star (as it doesn't depend on the Vaidya mass function) and is purely geometrical.
\item Also, it is quite a coincidence that classical free-gravity entropy at any given epoch is not extensive (as it also scales as area).
\end{enumerate}
Indeed, for two 3-hypersurfaces perpendicular to $u^a$, denoted by epochs $\tau_1$ and $\tau_2$ (where $\tau_2$ is in casual future of $\tau_1$), we have
\be
\delta\mathcal{S}_G\equiv\mathcal{S}_G(\tau_2)-\mathcal{S}_G(\tau_1)=-\frac14\delta A_\mathcal{B}\,,
\ee
where $\delta A_\mathcal{B}$ is the change in the surface area of the collapsing star between these epochs. As $\delta A_\mathcal{B}<0$ for a continual collapse, we have $\delta\mathcal{S}_G>0$ for the entire collapsing process, showing that the continual collapse is a thermodynamically viable spontaneous process with respect to the free-gravity thermodynamics.

{\it What happens classically when the surface of the collapsing star reaches the surface of infinite redshift according to the Vaidya observer?} Obviously, as seen by the Vaidya observer, the stellar surface freezes and $\delta A_\mathcal{B}\rightarrow 0$, $m(u)\rightarrow M$ (the black hole mass being $M$), $m(u)_{,u}\rightarrow 0$ (and hence the classical Vaidya radiation goes to zero). The renormalised free-gravitational entropy (over the value at infinity) then naturally attains the value of the Bekenstein-Hawking entropy $\mathcal{S}_G=\frac14A_{hor}$, with $\delta\mathcal{S}_G\rightarrow0$.

However, the entire process cannot stop here, as firstly, the spacetime still has non-trivial and non-zero curvature, and secondly, it will not be favourable according to the thermodynamics of free gravity, which would require $\delta\mathcal{S}_G>0$. Hence there must be some mechanism by which the horizon area reduces (that is, the BH evaporates). This is significant as {\it the emergence of quantum evaporation of BH is a necessity as demanded by the required monotonicity of the classical free-gravity entropy}.

To understand transparently, how the vacuum polarisation becomes important in the late stages of a continual collapse as the BH forms, we refer to the works of Leonard Parker  \cite{Par69}. Here it was clearly shown that the interior of the collapsing stellar fluid is in a time-varying gravitational field on account of the collapse process, leading to particle creation in a spacetime filled with an evolving fluid. This is exactly the mechanism of particle creation by a collapsing fluid forming a BH, as discussed by Hawking \cite{Hawking}, and Birrell and Davies \cite{BirDav84}. The key result of these calculations is: although the particle creation is primarily due to the collapsing fluid in the interior of the star, {\it the final outcome is independent of the nature and dynamics of the collapsing stellar interior} \cite{Ellis:2014jja}. This result goes hand in hand with our classical result \cite{suppl}.

At this point we are confronted by an important question: \emph{how would the quantum particles created by the collapsing matter disperse to infinity}? Following the pioneering work by Hiscock \cite{His81} and detailed calculations by Farley and D'Eath \cite{FarleyDEath}, we can claim that classical spacetimes which contain evaporating BH can be constructed using the Vaidya metric, thus ensuring the continuity of the collapsing stellar exterior in the regimes of both classical and quantum radiations.

In the comprehensive calculations presented in \cite{FarleyDEath}, quantum amplitudes arising in
gravitational collapse to a black hole were calculated following Feynman's method, by first rotating the Lorentzian time $T$ into the complex plane, that is, $T \rightarrow \vert T\vert\exp\{-i\alpha(u,r)\}$, (where   $0 < \alpha(u,r) \le \pi/2$), and  then solving the corresponding complex classical boundary-value problem to  compute the classical Lorentzian action $S_{class}$ and corresponding semi-classical quantum amplitude, proportional to $\exp(iS_{class})$. The final Lorentzian amplitude was then recovered by taking the limit $\alpha(u,r)\rightarrow 0_+$.

The key approach in these calculations was to perturb the background spacetime using first order scalar, vector and tensor potentials (corresponding to spin 0, spin 1 and spin 2  particles) and then to average out the cumulative energy momentum tensor over several wavelengths to produce a smooth, averaged and nearly spherically symmetric energy momentum tensor $T_{ab}$, that mimics a radially outgoing Vaidya null fluid over a suitably long time scale. The authors considered high frequency massless first order perturbations with the following potential ansatz
\ba
\psi^{(1)}=\sum_{l=0}^\infty\sum_{m=-l}^l\int_0^\infty d\omega [A_{lm\omega}\exp(i\alpha_w/\epsilon)+cc]\\
A_a^{(1)}=\sum_{l=1}^\infty\sum_{m=-l}^l\int_0^\infty d\omega [(A_a)_{lm\omega P}\exp(i\alpha_w/\epsilon)+cc]\\
h_{ab}^{(1)}=\sum_{l=2}^\infty\sum_{m=-l}^l\int_0^\infty d\omega [(A_{ab})_{lm\omega P}\exp(i\alpha_w/\epsilon)+cc]
\ea
where $P$ denotes the  orthogonal polarisations for gravitational waves, and also the standard polarisations of electromagnetic waves, and the free parameter $\epsilon$ keeps track of the magnitudes of different quantities in the highfrequency approximation. Defining $(k_a)_\omega=\nabla_a\alpha_\omega$, the Isaacson averaged energy momentum tensor was then obtained as
\be
\langle T_{ab}\rangle=\frac{2}{\epsilon^2}\sum_{slmP}c_s\int_0^\infty d\omega (k_a)_\omega(k_b)_\omega\vert A_{slm\omega P}\vert^2\;,
\ee
where $c_s=1, \frac{1}{4\pi},\frac{1}{32\pi}$ for $s=0,1,2$ respectively. This averaged energy momentum tensor has some interesting properties. Firstly, it transforms as a tensor with respect to the coordinate transformations in the $[u,e]$ plane. Secondly it obeys the conservation equation $\nabla^a\langle T_{ab}\rangle=0$, and finally we have  $\langle T^a_{~a}\rangle=0$ at the leading order, as it should be for the outgoing null radiation. Therefore, naturally the quantity $\vert A_{slm\omega P}\vert^2$ can be interpreted as the total intensity of the high frequency radiations of spin 0, spin 1, and spin 2  particles.

Casting this energy momentum tensor on a classical Vaidya background in the $ (u, r , \theta , \phi ) $ coordinate system, we then obtain from the field equations
\be
-\frac{1}{8\pi r^2}m(u)_{,u}=\sum_{slmP}c_s\int_0^\infty d\omega [ (k_u)_\omega ]^2 \vert A_{slm\omega P}\vert^2\;,
\ee
thus specifying the Vaidya mass function during the quantum radiation epochs.

Thus, we see that the role of the Vaidya observer remains intact, both during the classical radiation from the collapsing radiating star and the quantum radiation from the evaporating BH. The key difference between the classical and quantum regimes are: the geometrical and thermodynamic quantities related to this observer are sourced by the internal dynamics of the collapsing star in the classical phase, whereas in the quantum phase, these are sourced by the massless particles created near the horizon via vacuum polarisation (which are responsible for the evaporation of the BH). However, even in quantum case the expression for the total entropy calculated at any epoch (as given by equation (\ref{ent})) remains intact. In other words, the dynamics of the particle creation on the 3-volume of any epoch doesn't affect the total entropy at that epoch and it depends only on the horizon area. During the evaporation process, when the horizon of the BH shrinks down, the relation $\delta\mathcal{S}_G=-\frac14\delta A_{hor}$ holds exactly, giving $\delta\mathcal{S}_G>0$ for the entire process (till we are left with a flat Minkowski spacetime) again showing its thermodynamic viability and spontaneity.

We would like to emphasise an important point here. Although our classical calculations are done for the case of spherical symmetry, any non-spherical perturbation will evolve according to the background Regge-Wheeler potential \cite{Clarkson}
in the classical regime and will be radiated away via usual quasi normal modes of gravitational waves. This phenomenon is exactly replicated in the quantum regime by the spin 2 massless particles described above, which are generated by the tensor perturbations of the background spacetime. Therefore our mechanism is robust to any perturbations related to the various angular momenta of the collapsing star and evaporating BH.

To reiterate, we presented here an extremely robust geometrical and thermodynamical mechanism of free gravity that naturally extends from the classical radiating stellar collapse to a quantum evaporating BH, as seen by an external Vaidya observer. The key features of this mechanism are as follows:
\begin{enumerate}
\item The robustness, with respect to the internal dynamics of the collapsing star in the classical regime and the dynamics of vacuum polarisations in the quantum regime, remains intact throughout the process of collapse of an isolated star to a BH and then the evaporation of this BH at late times.
\item Further to this, the most remarkable aspect of free-gravitational entropy that emerged from our investigation is that it is non-extensive (classically) that has a natural transition to the Bekenstein-Hawking entropy of an evaporating BH. This definitely shows that the non extensiveness of the Bekenstein-Hawking entropy, which emerges from rigorous quantum calculations, does originate from the classical Riemannian geometry, that governs the classical free-gravitational entropy.
\item Finally, this entire mechanism remains valid for any non-spherical perturbation introduced by small angular momenta of the collapsing star and of the evaporating black hole.
\end{enumerate}

\begin{acknowledgments}
Portions of this work was done in IUCAA, India under the Associateship program, and in the University of KwaZulu-Natal (UKZN), South Africa. SG thanks IUCAA and UKZN. SK thanks Dr. Samarjit Chakraborty for initial inputs, and the West Bengal Government for research scholarship. RG is supported by the National Research Foundation (South Africa). The authors thank Prof. Naresh Dadhich for his comments.
\end{acknowledgments}

 \end{document}